\documentclass[conference,english]{IEEEtran}

\usepackage[T1]{fontenc}
\usepackage[style=ieee, doi=false, isbn=false, url=false]{biblatex}
\usepackage{hyperref}
\usepackage{bm,bbm}
\usepackage{amsmath,amssymb,mathtools}
\usepackage{bm}
\usepackage{stmaryrd}
\usepackage{microtype}
\usepackage[svgnames]{xcolor}
\usepackage{graphicx,tabularx,float}
\usepackage{babel,csquotes}
\usepackage{ifluatex}
\usepackage{comment}
\usepackage{adjustbox}
\usepackage{subcaption}
\usepackage{tikz}
\usetikzlibrary{positioning, shapes.multipart, arrows.meta, calc, decorations.pathreplacing}
\usepackage{verbatim}
\usepackage{enumitem}

\usepackage{algorithm,algorithmicx}
\usepackage[noend]{algpseudocode}

\makeatletter

\newtheorem{theorem}{Theorem}

\newtheorem{lemma}{Lemma}

\bmdefine{\bA}{A}
\bmdefine{\ba}{a}
\bmdefine{\bB}{B}
\bmdefine{\bb}{b}
\bmdefine{\bC}{C}
\bmdefine{\bc}{c}
\bmdefine{\bD}{D}
\bmdefine{\bd}{d}
\bmdefine{\bE}{E}
\bmdefine{\be}{e}
\bmdefine{\bF}{F}
\bmdefine{\bf}{f}
\bmdefine{\bG}{G}
\bmdefine{\bg}{g}
\bmdefine{\bH}{H}
\bmdefine{\bh}{h}
\bmdefine{\bI}{I}
\bmdefine{\bi}{i}
\bmdefine{\bJ}{J}
\bmdefine{\bj}{j}
\bmdefine{\bK}{K}
\bmdefine{\bk}{k}
\bmdefine{\bL}{L}
\bmdefine{\bl}{l}
\bmdefine{\bM}{M}
\bmdefine{\bmm}{m}
\bmdefine{\bN}{N}
\bmdefine{\bn}{n}
\bmdefine{\bO}{O}
\bmdefine{\bo}{o}
\bmdefine{\bP}{P}
\bmdefine{\bp}{p}
\bmdefine{\bQ}{Q}
\bmdefine{\bq}{q}
\bmdefine{\bR}{R}
\bmdefine{\br}{r}
\bmdefine{\bS}{S}
\bmdefine{\bs}{s}
\bmdefine{\bT}{T}
\bmdefine{\bt}{t}
\bmdefine{\bU}{U}
\bmdefine{\bu}{u}
\bmdefine{\bV}{V}
\bmdefine{\bv}{v}
\bmdefine{\bW}{W}
\bmdefine{\bw}{w}
\bmdefine{\bX}{X}
\bmdefine{\bx}{x}
\bmdefine{\bY}{Y}
\bmdefine{\by}{y}
\bmdefine{\bZ}{Z}
\bmdefine{\bz}{z}

\bmdefine{\balpha}{\alpha}
\bmdefine{\bbeta}{\beta}
\bmdefine{\bgamma}{\gamma}
\bmdefine{\bGamma}{\Gamma}
\bmdefine{\bdelta}{\delta}
\bmdefine{\bDelta}{\Delta}
\bmdefine{\bepsilon}{\epsilon}
\bmdefine{\bvarepsilon}{\varepsilon}
\bmdefine{\bzeta}{\zeta}
\bmdefine{\bmeta}{\eta}
\bmdefine{\btheta}{\theta}
\bmdefine{\bTheta}{\Theta}
\bmdefine{\biota}{\iota}
\bmdefine{\bkappa}{\kappa}
\bmdefine{\blambda}{\lambda}
\bmdefine{\bLambda}{\Lambda}
\bmdefine{\bmu}{\mu}
\bmdefine{\bnu}{\nu}
\bmdefine{\bpi}{\pi}
\bmdefine{\bPi}{\Pi}
\bmdefine{\brho}{\rho}
\bmdefine{\bsigma}{\sigma}
\bmdefine{\bSigma}{\Sigma}
\bmdefine{\btau}{\tau}
\bmdefine{\bupsilon}{\upsilon}
\bmdefine{\bUpsilon}{\Upsilon}
\bmdefine{\bphi}{\phi}
\bmdefine{\bPhi}{\Phi}
\bmdefine{\bchi}{\chi}
\bmdefine{\bpsi}{\psi}
\bmdefine{\bPsi}{\Psi}
\bmdefine{\bomega}{\omega}
\bmdefine{\bOmage}{\Omega}

\newcommand{\cL}{\mathcal{L}}

\newcommand{\bbC}{\mathbb{C}}

\DeclareMathOperator{\diag}{diag}

\DeclareMathOperator{\vectt}{vect}

\newcommand*{\herm}{{\mathsf{H}}}

\newcommand\norm[1]{\left\lVert #1 \right\rVert}

\makeatother
\usepackage{tikz}
\usepackage{amsmath}
\begin{document}

\title{Global Convergence of ESPRIT with Preconditioned First-Order Methods for Spike Deconvolution}

\IEEEoverridecommandlockouts

\hypersetup{
    pdftitle={Global Convergence of ESPRIT with Preconditioned First-Order Methods for Spike Deconvolution},
    pdfauthor={Joseph Gabet, Meghna Kalra, Maxime Ferreira Da Costa, and Kiryung Lee},
    pdfsubject={spike deconvolution, preconditioned gradient descent, ESPRIT, non-linear least squares, global convergence, super-resolution}
}

\author{
    \IEEEauthorblockN{
        Joseph Gabet\IEEEauthorrefmark{1}, 
        Meghna Kalra\IEEEauthorrefmark{2}, 
        Maxime Ferreira Da Costa\IEEEauthorrefmark{1}, 
        and Kiryung Lee\IEEEauthorrefmark{2}
    }
    \IEEEauthorblockA{
        \IEEEauthorrefmark{1}Laboratory of Signals and Systems, CentraleSupélec, Université Paris--Saclay, CNRS,
        Gif-sur-Yvette, France\\
    }
    \IEEEauthorblockA{
        \IEEEauthorrefmark{2}Ohio State University, Columbus, OH, USA\\
    }
  \thanks{
    JG and MF were supported in part by ANR funding ANR-24-CE48-3094, AID funding AID-2023639, and the Orange chair on ``Sustainable 6G'' held by CentraleSupélec. MK and KL were supported in part by NSF CAREER Award CCF-1943201.
    }
    \thanks{Authors' emails: joseph.gabet@centralesupelec.fr; kalra.42@osu.edu; maxime.ferreira@centralesupelec.fr; kiryung@ece.osu.edu}
}

\maketitle

\begin{abstract}
Spike deconvolution is the problem of recovering point sources from their convolution with a known point spread function, playing a fundamental role in many sensing and imaging applications. This paper proposes a novel approach combining ESPRIT with Preconditioned Gradient Descent (PGD) to estimate the amplitudes and locations of the point sources by a non-linear least squares. The preconditioning matrices are adaptively designed to account for variations in the learning process, ensuring a proven super-linear convergence rate.
We provide local convergence guarantees for PGD and performance analysis of ESPRIT reconstruction, leading to global convergence guarantees for our method in one-dimensional settings with multiple snapshots, demonstrating its robustness and effectiveness. Numerical simulations corroborate the performance of the proposed approach for spike deconvolution.
\end{abstract}


\section{Introduction}

Spike deconvolution, or super-resolution \cite{donoho1992superresolution}, addresses the recovery of discrete point sources from their convolution with a point spread function (PSF). It is a key challenge in many fields of applied science and engineering, such as radar, sonar, optical imaging, neuro-imaging, and communication systems \cite{potter2010sparsity, zhu2016super, zhu2012faster, berger2010sparse}. By counteracting the PSF's low-pass filtering effects, spike deconvolution aims to recover the original sources’ amplitudes and locations accurately.

Classical methods such as Prony’s algorithm, MUSIC \cite{schmidt1986multiple, liao2016music}, and ESPRIT \cite{roy1989esprit} rely on the algebraic structure of the problem to efficiently estimate the spikes. Yet, when the PSF is not an ideal low-pass filter, a prior equalization step is often required, with the effect of coloring and amplifying the noise statistics. Optimization-based approaches such as atomic norm minimization~\cite{decastro2012exact, tang2013compressed, candes2014towards, chi2020harnessing} stem a more flexible framework capable of adapting to arbitrary PSFs or missing observations. However, their reliance on semidefinite programming yields a significant computational overhead, limiting their scalability.

\subsection{Contributions, Prior Art, and Organization of the Paper}

In this paper, we study the problem of recovering point sources convolved by an \emph{arbitrary} known PSF from multiple snapshots of noisy Fourier-domain observations.
In recent work, the preconditioned gradient descent algorithm (PGD) was proposed as a scalable first-order method to estimate the source's parameters by minimizing a non-linear least squares program. By selecting an adequate diagonal preconditioning matrix, a lower estimate of the radius of the basin of attraction is obtained. The iterate sequence converges linearly in the absence of noise and for a single snapshot~\cite{ferreiradacosta2023LocalGeometry,gabet2024preconditioned}. Yet, the analysis relies on conservative bounds on the spectrum of the measurement operator, yielding a pessimistic dependency on the minimal resolvable source separation. Moreover, only the local geometry is considered, and the analysis lacks global convergence guarantees. Herein, PGD is enhanced to a full (non-diagonal) preconditioner, guaranteeing supra-linear convergence to the desired accuracy level in the presence of noise. Additionally, the minimal resolvable distance is explicated as a function of the PSF. Key to the analysis is the control of the extremal singular values of structured matrices through the Beurling--Selberg approximation theory~\cite{vaaler1985some,ferreira2023second}.

Due to the non-convex nature of the spike deconvolution problem, it is crucial to obtain a suitable initial estimate. We adopt the variant of ESPRIT by Swindlehurst and Gunther \cite{swindlehurst1999methods} for general PSFs. 
Their method was originally proposed for the blind case without the knowledge of the PSF. An intermediate step of deriving their main algorithm allows an ESPRIT algorithm that avoids unnecessary noise amplification. 
We present a rigorous perturbation analysis of this modified ESPRIT via the separation of spikes and attributes of the PSF. 
Although the modified ESPRIT provides the exact reconstruction in the noise-free or asymptotic regime under certain random noise scenarios, in practice, its performance under finitely many noisy snapshots is inferior to the direct optimization approach like PGD. Nevertheless, it provides a useful initialization for PGD backed by provable estimation guarantees. 
By combining the ESPRIT and PGD results, we guarantee that under sufficiently high SNR, the entire algorithm guarantees global convergence. 

Theoretical analyses of spike deconvolution have also been studied for different acquisition models \cite{bernstein2019deconvolution,traonmilin2020basins,traonmilin2024strong}. 
Furthermore, \cite{bresler1989resolution,kalra2024stable} showed that the ESPRIT algorithm resolves the common pulse locations in the blind case when the Fourier measurements of the pulse shape are pairwise similar for the two sub-arrays in ESPRIT. 

The rest of the paper is organized as follows. Our spike reconstruction problem is formulated in Section~\ref{sec:model}. 
Section~\ref{sec:pgd} presents the PGD algorithm, and its local convergence is established in Theorem~\ref{thm:conergence_pgd} as a function of the key problem's parameters, and under proviso of initializing close enough to the ground truth. Section~\ref{sec:esprit} introduces a modified ESPRIT method to adapt a non-ideal PSF. An upper bound on the statistical error is provided in  Theorem~\ref{thm:gen_case_deterministic}. Finally, Section~\ref{sec:experiments} discusses the global convergence properties of ESPRIT followed by PGD refinement, and numerical experiments corroborating our theory are conducted. A conclusion is drawn in Section~\ref{sec:conclusion}.

\subsection{Notation and Definitions}

Vectors $\bm{a}$ are denoted in bold letters and matrices $\bm{A}$ are in capital bold letters. The space of complex-valued matrices of size $u;v$ is denoted $\mathcal{M}_{u,v}(\mathbb{C})$. The transpose and Hermitian transpose of a matrix $\bm{A}$ are denoted by $\bA^\top$ and $\bA^\herm$, respectively. Its Moore-Penrose pseudo-inverse is denoted by $\bA^{\dagger}$. Its largest and smallest singular values are denoted by $\sigma_{\text{max}}(\bA)$ and $\sigma_{\text{min}}(\bA)$, respectively.  
We denote by $\mathbbm{1}_r$ and $\bm{0}_r$ respectively the all-one and all-zero vector in $\bbC^r$ and $\bO_{a,b}$ the all zeros matrix in $\mathcal{M}_{a,b}(\bbC)$ .
The Kronecker product is denoted $\bA \otimes \bB$.
For positive integers $a,b$, we denote by $\llbracket a;b\rrbracket$ the set of integers and then let $[a] = \llbracket 1;a\rrbracket$.

\subsection{Problem Formulation}
\label{sec:model}

We assume $L$ snapshots of observation (or channels) of a $r$-spikes signal sharing common support $\btau^{\star} = {[\tau^{\star}_1, \dots, \tau^{\star}_{r}]}^\top \in \mathbb{R}^{r}$ across the snapshots. The $r$ complex amplitudes of the spikes of the $\ell$-th snapshot are consigned in a vector $\bm{a}_\ell^\star \in \mathbb{C}^r$, which is further stacked in a matrix
$\bA^\star = \left[\bm{a}_1^\star, \cdots, \bm{a}_L^\star \right] \in \mathcal{M}_{r,L}(\mathbb{C})$
for convenience.
For any $(\bm{a}, \bm{\tau})$, we define the sparse complex Radon measure $\mu(\ba, \btau)
= \sum_{j=1}^r a_{j,\ell} \,\delta_{\tau_j}.$
After convolution with a known PSF $g \in L_1(\mathbb{R})$, the resulting $\ell$-th snapshot continuous domain signal $y_\ell$ reads
\begin{align*}
y_\ell(t) = (g \star \mu(\ba_\ell^\star, \bm{\tau}^\star))(t) + \xi_\ell(t),
\end{align*} where $\xi_\ell(t)$ is the noise in the continuous domain and $\star$ is the convolution product. 

Herein, we consider measurements $\by_\ell$ acquired in the Fourier domain, which is ubiquitous in signal processing and its applications. The continuous Fourier transform $\mathcal{F}(\cdot)$ of a Radon measure is defined by
\begin{equation*}
    \mathcal{F}(\mu)(f) = \int_{-\infty}^\infty  e^{-2 i \pi f \tau } d \mu(\tau),  \quad \forall f \in \mathbb{R}.
\end{equation*}
We fix ${N} = 2{n}+1$ as an odd number, and assume ${N}$ Fourier domain measurements taken uniformly over the frequency band $J_N = [-\tfrac{N}{2T},\tfrac{N}{2T}]$, so that the sampling set ${\Omega} \subset \mathbb{R}$ is given by ${\Omega} = \tfrac{1}{{T}}{\llbracket-{n}; {n}\rrbracket}$.
Letting  $\bm{G} = \diag \left([\mathcal{F}(g)(f)]_{f \in \Omega}\right)$, the discrete observation vector of the $\ell$-th snapshot $\by_\ell$ writes
\begin{align*}
     \bm{y}_\ell
     &= \bm{G} \Big( \,\sum_{j=1}^r a_{j,\ell}^\star e^{-2i\pi f \tau_j^\star} \Big)_{f \in \Omega}
     + \bm{z}_\ell,
\end{align*}
where $\bz_\ell = {[\mathcal{F}(\xi_\ell)]}_{f\in\Omega}$ is the noise after applying $\mathcal{F}$, which is assumed \emph{arbitrary}. In fine, the measurements can be compactly written in a matrix form  $\bY = \left[\by_1, \dots, \by_L \right]$ with
\begin{align}\label{eq:observation_model}
    \bY &= \bG \bV_{\btau^\star} \bA^\star + \bZ,
\end{align}
where $\bm{V}_{\bm{\tau}} \in \mathbb{C}^{N \times r}$ is the Vandermonde matrix with nodes $e^{-2i \pi \tau_j}$ and exponents in $\Omega$.

\subsection{Problem conditioning and critical metrics}\label{subsec:metrics}

We call $\Delta$ the \emph{minimal separation} between the true locations $\btau^\star$, which is defined as the smallest possible torus distance between two distinct positions. That is:
\begin{align*}
    \Delta \coloneqq \min_{j \neq j^\prime} \min_{p\in \mathbb{Z}}\vert \tau^\star_{j} - \tau^\star_{j^\prime} + pT\vert.
\end{align*} 
Let $E_g$ represent the band-limited energy of the PSF $g$ within the frequency band $J_N$ such that
\begin{align}
\label{eq:E_def}
E_g &\coloneqq \norm{ \widehat{g} \,  \mathbbm{1}_{J_N}}_{L_2}^2
\end{align}
where $\mathbbm{1}_{J_N}$ is the indicator function of the interval $J_N $. Furthermore, let $\rho_g$ measure the flatness of the power spectral density of $g$ within the interval $J_N$ and be defined as
\begin{equation}\label{eq:rho_def}
    \rho_g \coloneqq E_g^{-1} \cdot \norm{|\widehat{g}|^2 \, \mathbbm{1}_{J_N}}_{\mathrm{TV}}
\end{equation}
where the total variation norm $\norm{\cdot}_{\mathrm{TV}}$ of a measure $q$ is defined as
\(
    \norm{q}_{\mathrm{TV}} = \displaystyle \sup_{ \substack{h \in \mathcal{C}_0 \\ \norm{h}_{L_\infty} \leq 1}} \int_{-\infty}^{+\infty} h(f) \mathrm{d} q(f)
\).
Similarly, $E_{g^\prime}$, $\rho_{g'}$, and $\rho_{g''}$ are defined by \eqref{eq:E_def} and \eqref{eq:rho_def} when $g$ is substituted by $g'$ or $g''$.

\section{Local Convergence of Preconditioned Gradient Descent}\label{sec:pgd}
Given the measurement model~\eqref{eq:observation_model}, we introduce the quadratic loss $\cL$ between the observation and the explained signal by
\begin{align}\label{eq:noisy_loss_function}
    \mathcal{L}(\bA, \btau) &= \frac 1 2 
    \left\Vert 
    \bG \bV_\btau \bA - \bY
    \right\Vert_F^2,
\end{align}
and estimate the signal parameters $(\widehat{\bA}, \widehat{\btau})$ as the unconstrained minimizer of the loss~\eqref{eq:noisy_loss_function}. 
Since the loss~\eqref{eq:noisy_loss_function} is non-convex, multiple local minimizers may exist, and first-order optimization comes without global convergence guarantees. Moreover, the loss~\eqref{eq:noisy_loss_function} suffers from permutation ambiguities on the entries of $\btau$, implying the existence of multiple global minimizers.
Herein, we study the \emph{local geometry} of the loss~\eqref{eq:noisy_loss_function} in a small neighborhood of a ground-truth $(\bA^\star, \btau^\star)$, and characterize the width of the basin of attraction as a function of the parameters presented in Subsection~\ref{subsec:metrics}.

For optimization purpose, we vectorize the input parameters $(\bm{A}, \bm{\tau})$ in a variable $\bm{\theta} = {[ \bm{a}_1^\top, 
\cdots, \bm{a}_L^\top, \bm{\tau}^\top]}^\top \in \mathbb{C}^{r(L+1)}$. 
We present the expression of the gradient of the loss~\eqref{eq:noisy_loss_function} with respect to amplitudes and positions. For that purpose, we introduce
\[
\bM_{\bm{A}} = \left[\begin{array}{c|c}
    \bm{I}_{Lr} & \bm{O}_{Lr,r} \\\hline
    \bm{O}_{Lr,Lr} & \diag(\bm{a}_{1}) \\
    \vdots &  \vdots\\
    \bm{O}_{Lr,Lr} & \diag(\bm{a}_{L})
    \end{array}\right]
 \in \mathcal{M}_{2Lr,(L+1)r}(\mathbb{C}),
\]
and let $\bLambda$ the diagonal matrix with $[\bm{\Lambda}]_{u,u} = -2 i \pi T^{-1} f_u$, and
\begin{align*}
\bW_\btau &= \left[\bI_L \otimes \bV_\btau \,|\, \bI_L \otimes \bm{\Lambda} \bV_\btau \right] \,\ \in \mathcal{M}_{LN,2Lr}(\mathbb{C}).
\end{align*} 
Furthermore, we write the shorthand $\bM^\star = \bM_{\bA^\star}$ and $\bW^\star = \bW_{\btau^\star}$. 

Then the gradient of $\cL$ in Equation~\eqref{eq:noisy_loss_function} is written as
\begin{multline} \label{eq:gradient}
    \nabla \mathcal{L}(\bm{\theta}) = \bm{M}_{\bA}^\herm \bm{W}_{\btau}^\herm \Bigg( \bm{W}_{\btau} \bm{M}^\star
    \begin{bmatrix} \vectt(\bm{A)} \\ \bm{0}_r \end{bmatrix} \\
    - \bm{W}^\star \bm{M}^\star \begin{bmatrix} \vectt(\bA^\star)  \\ \bm{0}_r \end{bmatrix} + \vectt(\bm{Z}) \Bigg).
\end{multline}
Given the heterogeneous nature of vector $\bm{\theta}$, which contains both locations and amplitudes, the loss~\eqref{eq:noisy_loss_function} and its gradient~\eqref{eq:gradient} are in general ill-conditioned, especially under a large dynamic range. Therefore, we rely on \emph{preconditioning}~\cite{ferreiradacosta2023LocalGeometry}---a quasi-Newton method---to adapt the direction of descent to the local landscape of the cost function. This is achieved by multiplying at step $k$ the gradient by a matrix $\bP_k$, which depends on the current estimate $\bm{\theta}_k$.
We select the preconditioner
\begin{equation}\label{eq:Pk-def}
    \bm{P}_k = \left( \bm{M}_{\bA_k}^\herm \bm{W}_{\btau_k}^\herm \bm{W}_{\btau_k} \bm{M}_{\bA_k}\right)^{-1}.
\end{equation}
The PGD algorithm is detailed in Algorithm~\ref{alg:PGD}. 
\begin{algorithm}[t]
\caption{Preconditioned Gradient Descent (PGD)}
\begin{algorithmic}[1]
\State \textbf{input} \((\bA_0, \bm{\tau}_0)\); $k \gets 0$.
\While{stopping criterion is not met}
\State Compute $\bP_k$ as in~\eqref{eq:Pk-def}
\State  \(\btheta_{k+1} \gets \bm{\theta}_k - \bP_k  \mathcal{L}(\bm{\theta}_k) \).
\State \(k \gets k+1\)
\EndWhile
\State \textbf{return} \(\bm{\theta}_k\)
\end{algorithmic}\label{alg:PGD}
\end{algorithm}

To study the convergence of the PGD sequence, we introduce the $\bm{u}_k, \bm{u}^\star \in \mathbb{C}^r$ the row $\ell_2$-norm vector of the matrices $\bm{A}_k$ and $\bm{A}^\star$, respectively. That is
\(    u_j^{\star} = \sqrt{\sum_{\ell=1}^L {{|a_{j,\ell}^{\star}}|}^2}
\) for all $j\in [N]$. Then, the \emph{weighted} error $\eta_k$ between the ground-truth parameter and the $k$-th iterate is selected as
\begin{align}\label{eq:weighted_error}
    \eta_k = \sqrt{E_g \sum_{j=1}^r \sum_{\ell = 1}^L\frac{\left\vert a^\star_{j,\ell}\right\vert^2}{u{^\star_j}^4} 
     \left\vert a^{(k)}_{j,\ell} - a^\star_{j,\ell} \right\vert^2 + E_{g^\prime} \sum_{j=1}^r \left\vert \tau^{(k)}_{j} - \tau^\star_j \right\vert^2}.
\end{align}
The weighting~\eqref{eq:weighted_error} fairly captures the error on the amplitudes in the event of a large dynamic range and enforces invariance of amplitude and location errors both with respect to the number of measurement $N$ and the number of snapshots $L$.

\begin{theorem}[Local convergence of PGD]\label{thm:conergence_pgd}
    Suppose $\Delta > \frac{2}{3}\rho_{g'}$ and let the quantities
    \begin{align*}
    \alpha &= 1 + \frac{u^\star_{\max}}{u^\star_{\min}} \frac{\sqrt{E_{g^{\prime\prime}}}}{E_{g^\prime}} \sqrt{\frac{1 + \frac{1}{2} \rho_{g''} \Delta^{-1}}{1-\tfrac{2}{3}\rho_{g'} \Delta^{-1}}} \\
    \beta &= \frac{1}{\sqrt{T E_{g'}(1-\tfrac{2}{3}\rho_{g'} \Delta^{-1})}}.
    \end{align*}
    If $4(\alpha+1)\beta \left\vert u_{\min}^\star \right\vert^{-1} \left\Vert \bm{Z} \right\Vert_F \leq 1$ and the initialization $\bm{\theta}_0$ has a weighted error $\eta_0$ satisfying
    \begin{align*}
        \eta_0 < \frac{1 + \sqrt{1- 4(\alpha+1)\beta  {u_{\min}^\star}^{-1} \left\Vert \bm{Z} \right\Vert_F}}{2 (\alpha+1)},
    \end{align*}
    then the error sequence ${\{\eta_k\}}_k$ satisfies
    \begin{equation} \label{eq:cv}
        \limsup_{k\to \infty} \eta_k \leq  \frac{1 - \sqrt{1- 4(\alpha+1)\beta  {u_{\min}^\star}^{-1} \left\Vert \bm{Z} \right\Vert_F}}{2 (\alpha+1)} \eqqcolon \gamma_\infty.
    \end{equation}
    Furthermore, ${\{\eta_k\}}_k$ converges super-linearly into $[0,\gamma_\infty]$.
\end{theorem}

\begin{figure}
\center
\begin{tikzpicture}[line cap=round, line join=round, xscale=11, yscale=9]

  \draw[->,thick] (-0.02,0) -- (0.4,0) node[right] {$\eta$};
  \draw[->,thick] (0,-0.02) -- (0,0.35) node[above] {};

  \draw[red,thick,dashed,samples=200,domain=-0.02:0.35]
       plot(\x,\x);

  \draw[Green,thick,samples=200,domain=-0.02:0.33]
       plot(\x,{(2*(\x)^2 + 0.03)/(1-\x)});

  \coordinate (Xinf)  at (0.033333, 0.033333);
  \coordinate (Xzero) at (0.3, 0.3);

  \fill[red] (Xinf) circle(0.3pt);
  \fill[red] (Xzero) circle(0.3pt);

  \draw[dotted,black] (Xinf) -- (Xinf|-0,0);
  
  \coordinate (eta0) at (0.25, 0);
  \node[black,below] at (eta0) {$\gamma_0$};
  \coordinate (f1eta0) at (0.25, 0.206667);
  \draw[dotted,blue] (eta0) -- (f1eta0);
  \coordinate (eta1) at (0.206667,0.206667);
  \draw[solid,blue] (f1eta0) -- (eta1);
  \coordinate (f1eta1) at (0.206667, 0.145491);
  \draw[solid,blue] (eta1) -- (f1eta1);
  \node[black,below] at (eta1|-0,0){$\gamma_1$};
  \draw[dotted,blue] (eta1) -- (eta1|-0,0);
  \coordinate (eta2) at (0.145491,0.145491);
  \node[black,below] at (eta2|-0,0){$\gamma_2$};
  \draw[dotted,blue] (eta2) -- (eta2|-0,0);
  \draw[solid,blue] (f1eta1) -- (eta2);
  \coordinate (f1eta2) at (0.145491, 0.084651);
  \draw[solid,blue] (eta2) -- (f1eta2);
  \coordinate (eta3) at (0.084651,0.084651);
  \node[black,below] at (eta3|-0,0){$\gamma_3$};
  \draw[dotted,blue] (eta3) -- (eta3|-0,0);
  \draw[solid,blue] (f1eta2) -- (eta3);
  \coordinate (f1eta3) at (0.084651, 0.048431);
  \draw[solid,blue] (eta3) -- (f1eta3);
  \coordinate (eta4) at (0.048431,0.048431);
  \draw[solid,blue] (f1eta3) -- (eta4);

  \node[black,below] at (Xinf|-0,0) {$\gamma_\infty$};

\end{tikzpicture}
\vspace{-24pt}
\caption{Illustration of the convergence of PGD. The weighted error sequence satisfies $\eta_{k+1} \leq f_1(\eta_k)$ where $f_1(\eta) = \frac{\alpha \eta^2 + \beta |u^\star_{\min}|^{-1}\|\bZ\|_F}{1-\eta}$ (solid;green). When the sequence $(\eta_k)_k$ starts between the two fixed points of $f_1(\eta)$ and $f_2(\eta) = \eta$ (dashed;red), its upper bound $(\gamma_k)_k$ converges to the left fixed point $\gamma_\infty$ super linearly.}
\end{figure}
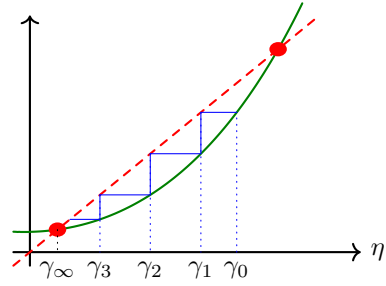

In the absence of noise, Theorem~\ref{thm:conergence_pgd} indicates the PGD recovers the ground truth parameters $\btheta^\star$ under proviso of a good enough initialization. Furthermore, the choice of the preconditioner~\eqref{eq:Pk-def} enables quadratic convergence towards the ground truth.
When the columns of $\bA^\star$ follow an isotropic distribution and $L \gg r$, the ratio $\frac{u^\star_{\max}}{u^\star_{\min}}$ concentrates with high probability around $1$, and the residual error is robust to the presence of a few ill-conditioned entries of the matrix $\bA^\star$.   
For Equation~\eqref{eq:cv}, in the high-SNR regime, $\gamma_\infty \simeq \beta {u_{\min}^\star}^{-1} \| \bZ \|_F$, and the noise amplification factor depends on the flatness measure of the power spectral density of the PSF and of the separation between the spikes.

\section{ESPRIT Initialization Using Known PSF}\label{sec:esprit}
We propose using ESPRIT \cite{roy1989esprit} to provide an initialization to PGD, given the non-convex nature of the problem. Specifically, we adopt the variant of ESPRIT by Swindlehurst and Gunther \cite{swindlehurst1999methods}
which utilizes the known PSF. 
Their method first finds $\widehat{\bm U} \in \mathbb{C}^{N \times r}$ that spans an estimate of the column space of $\bG \bV_{\btau}$ from the noisy observations $\bm{Y}$. 
It constructs two sub-matrices $\widehat{\bm U}_1 \coloneqq \bm{\Pi}_1 \widehat{\bm U}$ and $ \widehat{\bm U}_2 \coloneqq \bm{\Pi}_2 \widehat{\bm U}$ of $\widehat{\bm U}$ where $\bm{\Pi}_1:= \left[\bm{I}_{N-1} \,|\, \bm{0}_{(N-1)\times 1} \right]$ and $\bm{\Pi}_2 := \left[ \bm{0}_{(N-1)\times 1} \,|\, \bm{I}_{N-1} \right]$. 
Then, the locations can be estimated (up to a permutation ambiguity) as
\begin{equation}
\label{eq:m_esprit}
\widehat{\tau}_k = - \frac{T}{2\pi} \arg(\lambda_k(\widehat{\bm U}_1^\dagger \bm{G}_1\bm{G}_2^{-1}\widehat{\bm U}_2)), \quad k \in [r] 
\end{equation}
with $\bG_1 \coloneqq \bPi_1 \bG \bPi_1^\top$ and $\bG_2 \coloneqq \bPi_2 \bG \bPi_2^\top$, where $\mathrm{arg}(z)$ extracts the argument of the input $z \in \mathbb{C}$. 
Below we present the extension of the theoretical analysis of the original ESPRIT \cite{roy1989esprit} in the Dirac PSF case \cite{li2022stability} to the variant by \eqref{eq:m_esprit} for non-Dirac PSFs.
The accuracy of the estimated locations $\widehat{\btau}$ are
compared to the ground-truth $\bm\tau^\star$ via the
\emph{matching distance metric} defined by
\begin{align*}
\mathrm{md}(\widehat{\btau} ; \btau^\star) = \min_{\pi \in \mathfrak{S}_r} \max_{k \in [r]}  \vert \tau^\star_k - \widehat{\tau}_{\pi(k)} \vert
\end{align*}
where $\mathfrak{S}_r$ denotes the set of all permutations of $[r]$.
\begin{theorem}
\label{thm:gen_case_deterministic}
Let $N \geq r+1$. Let $\bU \in \mathbb{C}^{N \times r}$ span the column space of $\bG \bV_{\btau^\star}$ and
satisfy $\bU^\mathsf{H} \bU = \bI_r$.
Define 
\begin{align*}
    \varrho_g := \max_{m \in [N-1]}\frac{(\bm{G}_2)_{m,m}}{(\bm{G}_1)_{m,m}}. 
\end{align*}
Suppose that $\bm A$ has full row rank and
\begin{align}
\label{eq:prop_cond}
\mathrm{dist}(\bm{\widehat{U}},\bm U)
< \frac{1}{2\sqrt{2}} \sqrt{1 - \frac{\|\bG\|^2 r}{TE_{g} \left(1 - \frac{1}{2} \rho_g \Delta^{-1} \right) }} 
\end{align}
where $\mathrm{dist}(\bm{\widehat{U}},{\bm{U}}) :=
\|\widehat{\bm U} \widehat{\bm U}^\mathsf{H} - \bm{U} \bm{U}^\mathsf{H}\|$ corresponds to the sine of the largest principal angle between the subspaces spanned by $\bm{U}$ and $\widehat{\bm U}$. 
Then the estimate $\widehat{\btau}$ by ESPRIT satisfies
\begin{align}
\label{eq:md_tau_deterministic}
\mathrm{md}(\widehat{\btau} ; \btau^\star)  
& \lesssim T \varrho_g
\left(1 - \frac{\|\bG\|^2 r}{TE_{g} \left(1 - \frac{1}{2} \rho_g \Delta^{-1} \right)}\right)^{-1} \mathrm{dist}(\bm{\widehat{U}},\bm U).
\end{align}
\end{theorem}
The subspace estimation error is upper-bounded by the Davis--Kahan Theorem~\cite{davis1970rotation} as
\begin{align} \label{eq:daviskahan_fs2}
\mathrm{dist}(\bm{\widehat{U}},\bm{U}) 
\leq \frac{2 \min_c \norm{\bG \bV_{\btau} \bZ^\mathsf{H} + \bZ \bV_{\btau}^\mathsf{H} \bG^\mathsf{H} + \bZ \bZ^\mathsf{H} - c \bI_N}}{TE_{g} \left(1 - \frac{1}{2} \rho_g \Delta^{-1} \right)}
\end{align}
Therefore, when the noise level is sufficiently low, the condition in \eqref{eq:prop_cond} to invoke Theorem~\ref{thm:gen_case_deterministic} is satisfied. 
Note that $T E_g$ dominates $r \|\bG\|^2$ for narrow PSFs. For example, if $g$ is Dirac, then $T E_g = N$ dominates $r \|\bG\|^2 = r$. 
The error amplification factor $\varrho_g$ reflects the relative change of two consecutive Fourier magnitudes of $g$, which does not depend crucially on the dynamic range of $\bG$. 
The propagation of the ESPRIT estimate of $\btau$ to the subsequence least squares estimate of amplitudes is upper-bounded as shown in the following lemma. 
\begin{lemma}\label{lem:bound_a}
If $\mathrm{md}(\widehat{\btau} ; \btau^\star) \leq \delta$, then the least squares estimate $\widehat{\bA}$ given $\widehat{\btau}$ satisfies
\begin{align*}
 \frac{{\|\bA^\star - \widehat{\bA}\|}_F}{{\|\bA^\star\|}_F} &\leq \delta \sqrt{\frac{E_{g^\prime}}{E_g} \frac{1 + \frac{2}{3} \rho_{g^\prime} {(\Delta-2\delta)}^{-1}}{1 - \frac{1}{2}\rho_g {(\Delta-2\delta)}^{-1}}}.
\end{align*}

\end{lemma}
Lemma~\ref{lem:bound_a} implies that when the initial error in locations by ESPRIT $\delta< \Delta/2 - \max\left({\rho_{g'},\rho_g}\right)$, then the normalized error in recovering the amplitudes ends up being within a constant factor of the initial error by ESPRIT.

\section{Global Convergence of PGD Initialized with ESPRIT and Numerical Results}\label{sec:experiments}

Equations~\eqref{eq:md_tau_deterministic} and \eqref{eq:daviskahan_fs2} from Theorem~\ref{thm:gen_case_deterministic} show that if the noise level is sufficiently low, then ESPRIT reconstructs $\widehat{\btau}$ within any desired error level. Therefore, by Lemma~\ref{lem:bound_a}, the least square estimation error on $\widehat{\bA}$ can also be made sufficiently small for a large enough SNR. Consequently, initializing PGD with ESPRIT unsure global convergence of the non-convex optimization algorithm at a super-linear rate, under the separation condition of  Theorem~\ref{thm:conergence_pgd}.

\begin{figure}[t]
    \centering
    \begin{subfigure}[t]{0.48\columnwidth}
        \centering
        \includegraphics[width=\linewidth]{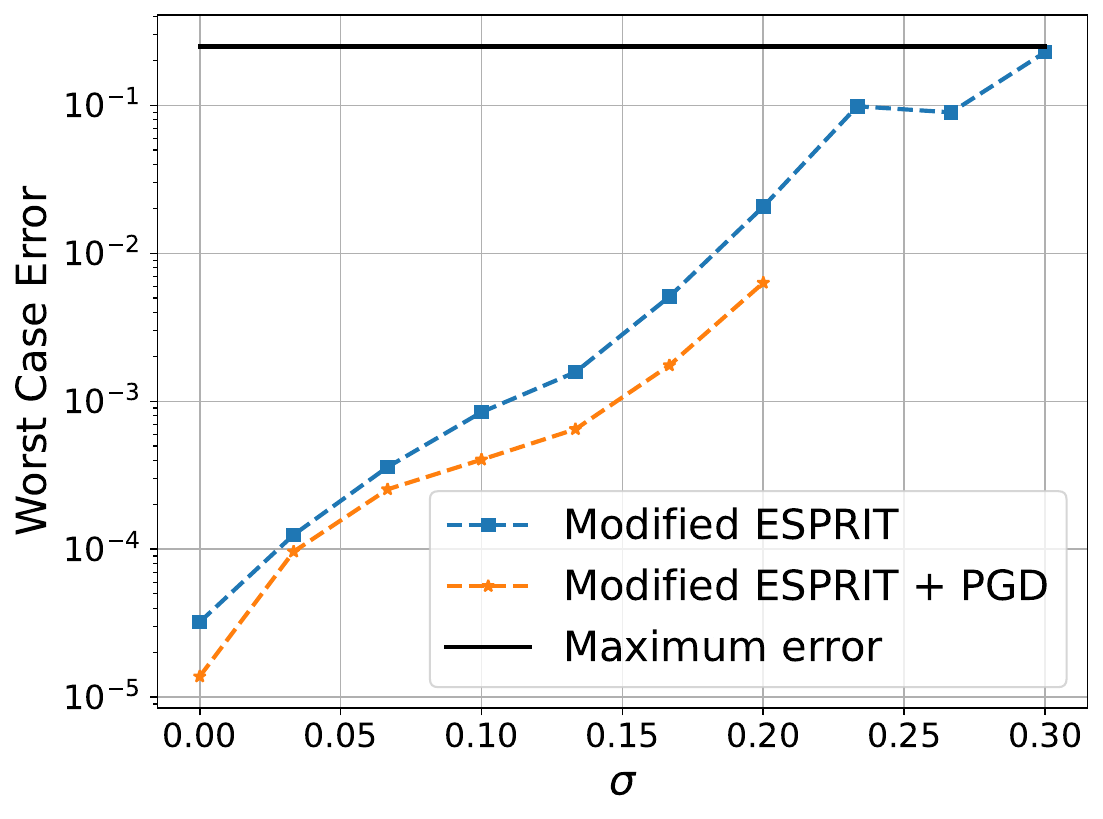}
        \caption{\centering Error with respect to variance 
        for Gaussian PSF (SNR=25dB)}
        \label{fig:error_vs_sigma}
    \end{subfigure}
    \hfill
    \begin{subfigure}[t]{0.48\columnwidth}
    \centering\includegraphics[width=\linewidth]{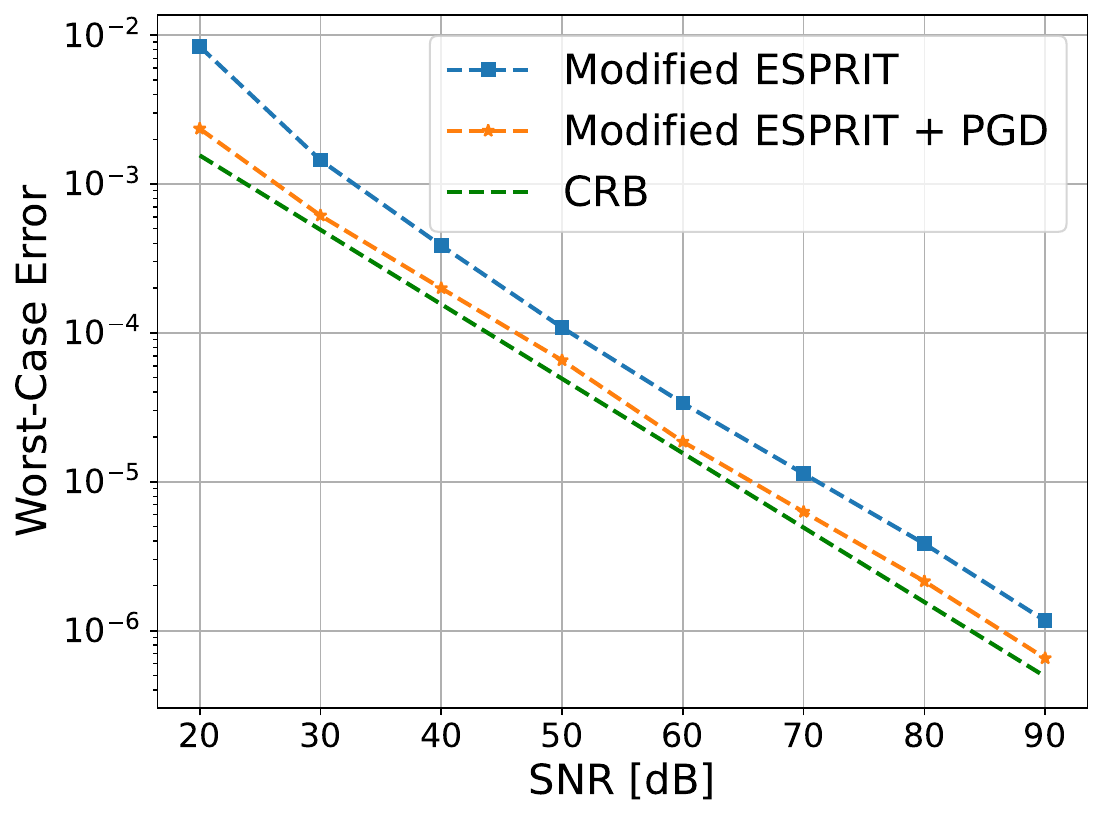}
    \caption{Error with respect to SNR \\ \centering ($\sigma = 0.15$)}
    \label{fig:error_vs_snr}
    \end{subfigure}
    \caption{Performance of ESPRIT and ESPRIT plus PGD.} 
    \label{fig:comparison}
\end{figure}

Figure~\ref{fig:error_vs_sigma} compares the worst-case error on the spike location $\btau^\star$ produced by modified ESPRIT and modified ESPRIT followed by PGD as a function of the width of the PSF, under a constant SNR. When the pulse is narrow (\emph{i.e.} small $\sigma$), the PGD refinements improve the estimation error. However, when the PSF width crosses a threshold, ESPRIT reconstruction error is nearly maximum, falling out of the convergence range of PGD predicted by Theorem~\ref{thm:conergence_pgd}.
Figure~\ref{fig:error_vs_snr} compares the same error with respect to SNR for a fixed PSF. While ESPRIT's error decays consistently as the SNR increases, the PGD refinements further enhance the location estimate.

\section{Conclusion}\label{sec:conclusion}

We have proposed a novel algorithm combining ESPRIT and a preconditioned gradient descent (PGD) to tackle a challenging non-convex optimization problem in spike deconvolution within the Fourier domain. This approach effectively identifies both the amplitudes and locations of sources, with theoretical guarantees of global convergence under mild assumptions, involving the separation of the sources. Our analysis, presented in Theorem~\ref{thm:conergence_pgd} and~\ref{thm:gen_case_deterministic}, leverages existing results on ESPRIT and introduces new, tighter bounds for the convergence of PGD.

While the proposed method demonstrates promising results, certain limitations remain. Notably, we assume prior knowledge of the number of sources, which may not always be available in practice and could require separate estimation algorithms. Furthermore, our framework currently considers a one-dimensional setting with a known point spread function (PSF). Extending the study to multidimensional cases or exploring guarantees in blind spike deconvolution scenarios represents a natural direction for future work. 

\newpage
\IEEEtriggeratref{4}
\renewcommand*{\bibfont}{\footnotesize}
\printbibliography

\end{document}